\documentclass[authoryear,preprint,12pt]{elsarticle}

\usepackage{amssymb}

\usepackage{subcaption}
\usepackage{wrapfig,caption}
\usepackage{multirow}
\usepackage{amsmath}
\usepackage{upgreek}
\usepackage{hyperref}
\usepackage{fancyhdr}
\usepackage{enumitem}
\usepackage{color}
\usepackage{array}
\usepackage{amstext}
\usepackage{amsmath} 
\usepackage{amssymb}
\usepackage[usenames,dvipsnames,svgnames,table]{xcolor}
\usepackage{psfrag}
\usepackage{array}
\usepackage{threeparttable}
\usepackage{dcolumn}
\newcolumntype{d}{D{.}{.}{-1}}
\usepackage{enumitem}
\usepackage{titlesec}
\usepackage{float}
\usepackage{comment}

\newcommand\Fr{\mbox{\text{Fr}}}  

\usepackage{xcolor}
\usepackage{tikz}
\usepackage{pgfplots}
\usepackage{bm}

\pgfplotsset{
        layers/my layer set/.define layer set={
            background,
            main,
            foreground
        }{
        },
        set layers=my layer set,
    }

\definecolor{findOptimalPartition}{HTML}{D7191C}
\definecolor{storeClusterComponent}{HTML}{FDAE61}
\definecolor{dbscan}{HTML}{ABDDA4}
\definecolor{constructCluster}{HTML}{2B83BA}

\definecolor{ColorOne}{HTML}{61BB46}
\definecolor{ColorTwo}{HTML}{FDB827}
\definecolor{ColorThree}{HTML}{F5821F}
\definecolor{ColorFour}{HTML}{E03A3E}
\definecolor{ColorFive}{HTML}{963D97}
\definecolor{ColorSix}{HTML}{009DDC}
\definecolor{Invisible}{HTML}{FFFFFF}

\usepackage{tikz}
\usepackage{blindtext}
\usetikzlibrary{plotmarks}

\usepackage{pgfplots}
\pgfplotsset{compat=newest}

\usetikzlibrary{plotmarks}
\usepackage{array,booktabs}

\usepackage{scalerel}
\usepackage{stackengine,wasysym}

\journal{Acta Mechanica Sinica}

\begin{document}

\begin{frontmatter}



\title{Multiphase turbulence modeling using sparse regression and gene expression programming}


\author[label1]{Sarah Beetham}
\author[label1,label2]{Jesse Capecelatro}
\address[label1]{University of Michigan, Department of Mechanical Engineering}
\address[label2]{University of Michigan, Department of Aerospace Engineering}

\begin{abstract}
In recent years, there has been an explosion of machine learning techniques for turbulence closure modeling, though many rely on augmenting existing models. While this has proven successful in single-phase flows, it breaks down for multiphase flows, particularly when the system dynamics are controlled by two-way coupling between the phases. 
In this work, we propose an approach that blends sparse regression and gene expression programming (GEP) to generate closed-form algebraic models from simulation data. Sparse regression is used to determine a minimum set of functional groups required to capture the physics and GEP is used to automate the formulation of the coefficients and dependencies on operating conditions. The framework is demonstrated on a canonical gas--solid flow in which two-way coupling generates and sustains fluid-phase turbulence.  

\end{abstract}

\end{frontmatter}

\section{Introduction}
\label{sec:intro} 
In the last decade, data-driven approaches have become the predominant tool for developing turbulence models~\citep{duraisamy2019turbulence}. Of these approaches, Neural Networks (NNs) have gained a considerable amount of traction \citep{ML_2015Tracey, ML_2002Milano, ML_2010Lu, ML_2012Rajabi, ML_2014Duraisamy_transition, ML_2014Duraisamy_new, ML_2016Ma, Ling2016, Bode2019, Liu2019}. In contrast, a relative minority of approaches have elected to pursue strategies that enable a compact, algebraic closure. Formulating models in this way has several important properties including increased interpretability, ease of dissemination and straightforward integration into existing solvers. These techniques generally fall into two categories, (1) symbolic regression and (2) gene expression programming. 

In the case of sparse regression, \citet{ML_2016Brunton} developed a strategy based on sparse regression that identifies the underlying functional form of the nonlinear physics by optimizing a coefficient matrix that acts upon a matrix of trial functions. This construct has the important benefit of including the user in the modeling process through selection of the trial functions. \citet{schmelzer2020discovery,Beetham2020} recently extended the sparse identification framework of \citet{ML_2016Brunton} to infer algebraic stress models for the closure of the Reynolds-averaged Navier--Stokes (RANS) equations. In \citet{schmelzer2020discovery}, the models are written as tensor polynomials and built from a library of candidate functions. In \citet{Beetham2020}, Galilean invariance of the resulting models are guaranteed through thoughtful tailoring of the feature space.

Gene expression programming (GEP), a data-driven technique inspired by the Darwinian concept of survival-of-the-fittest, heuristically evolves symbolic models until error is reduced beyond a threshold. In recent years, this strategy has gained attention in the turbulence modeling. For example, GEP has demonstrated success in the contexts of modeling large eddy simulation (LES) subgrid scale closures \citep{reissmann2021application}, boundary layer theory \citep{dominique2021inferring}, turbulent pipe flow \citep{samadianfard2012gene} and informing RANS closures \citep{weatheritt2016novel, zhao2020rans}. 

While these data-driven techniques have been increasingly utilized for modeling single-phase turbulence, their application to multiphase turbulence modeling is still relatively uncharted. Despite this, multiphase flows present a rich and diverse class of problems for which machine learning can prove useful. Due to the large parameter space frequently attributed to such flows, traditional modeling techniques have historically failed, especially beyond dilute regimes, where models extended from single-phase turbulence break down \cite[see, e.g.,][and discussion therein]{fox2014}. This divergence from single-phase turbulence theory can be attributed to the fact that at moderate volume fractions, particles generate turbulent kinetic energy (TKE) at the smallest scales. This is the direct antithesis to the classical notion of the turbulent energy cascade.  
Additionally, numerous practical applications span regimes from dilute to dense particle loadings, motivating the need for models that are accurate across regimes. This motivates the need for methodologies capable of formulating closures `from scratch,' rather than augmenting existing models. These challenges, along with the societal importance and pervasiveness of these flows, make them excellent candidates for improvements in data-driven modeling techniques. 

In this work, we propose a blending modeling approach which combines the strengths of both sparse regression and GEP to inform multiphase turbulence closures in a way that leverages the physical knowledge of the modeler and automates the determination of model components for which physical insight is not obvious or does not exist. To demonstrate the utility of such an approach, we present a simple configuration in which strong two-way coupling between fluid and particles generates and sustains turbulence. This configuration has been discussed extensively in prior work \cite[see, e.g.,][for more exhaustive details]{capecelatro2014cit,capecelatro2015,Beetham2020} and serves as a case study here.



\section{Methodology}
\label{sec:method} 
It is well established \citep{PopeText} that any tensor quantity, $\mathcal{D}_{ij}$, can be exactly described by an infinite sum given as
\begin{equation}
    \mathcal{D}_{ij} = \sum_{n=1}^{\infty} \beta^{(n)} \mathcal{T}_{ij}^{(n)}, \label{eq:linear}
\end{equation}
where $\beta^{(n)}$ represents the $n$-th coefficient associated with a corresponding basis tensor, $\mathcal{T}_{ij}^{(n)}$. The coefficients may range in complexity from constants to nonlinear functions of the principal invariants of the tensor bases.  For many configurations, this infinite sum can reduced to a finite sum by leveraging the Cayley--Hamilton theorem. This results in a reduced set of tensors termed a minimal invariant basis \citep[e.g.,][]{Spencer1958}. Using knowledge of the system physics, a minimal invariant basis can be derived. Once this basis is established, modeling can be broken into two tasks: (1) \emph{Which of the basis tensors are most important for capturing the physics at play?} and (2) \emph{How do the coefficients depend on principal invariants or system parameters?} 

Sparse regression has been shown to be successful at addressing the first task \citep{Beetham2020, Beetham2021} and works well for the second task when \emph{constant} coefficients are sufficient. However, when the system has a complex and large parameter space, as is the case for multiphase turbulence, constant coefficients are no longer sufficient for capturing physics across scales. In this situation, sparse regression is not an efficient method for determining the form of the coefficients and requires the modeler to supply all potential test functions to the algorithm manually. While this has important benefits for embedding physics-based reasoning and properties into the resultant model (e.g., form invariance), it implies a tedious, `guess-and-check' exercise if physics-based arguments can no longer be used to supply test functions. In the present method, we propose to offload this work to a gene expression algorithm when naivity in functional form is unavoidable. This effectively automates the process of evaluating trial functions for the coefficients, while preserving the benefits of using sparse regression to inform the tensorial building blocks of the model.    

\begin{figure}[ht]
    \centering
    \includegraphics[width=\textwidth]{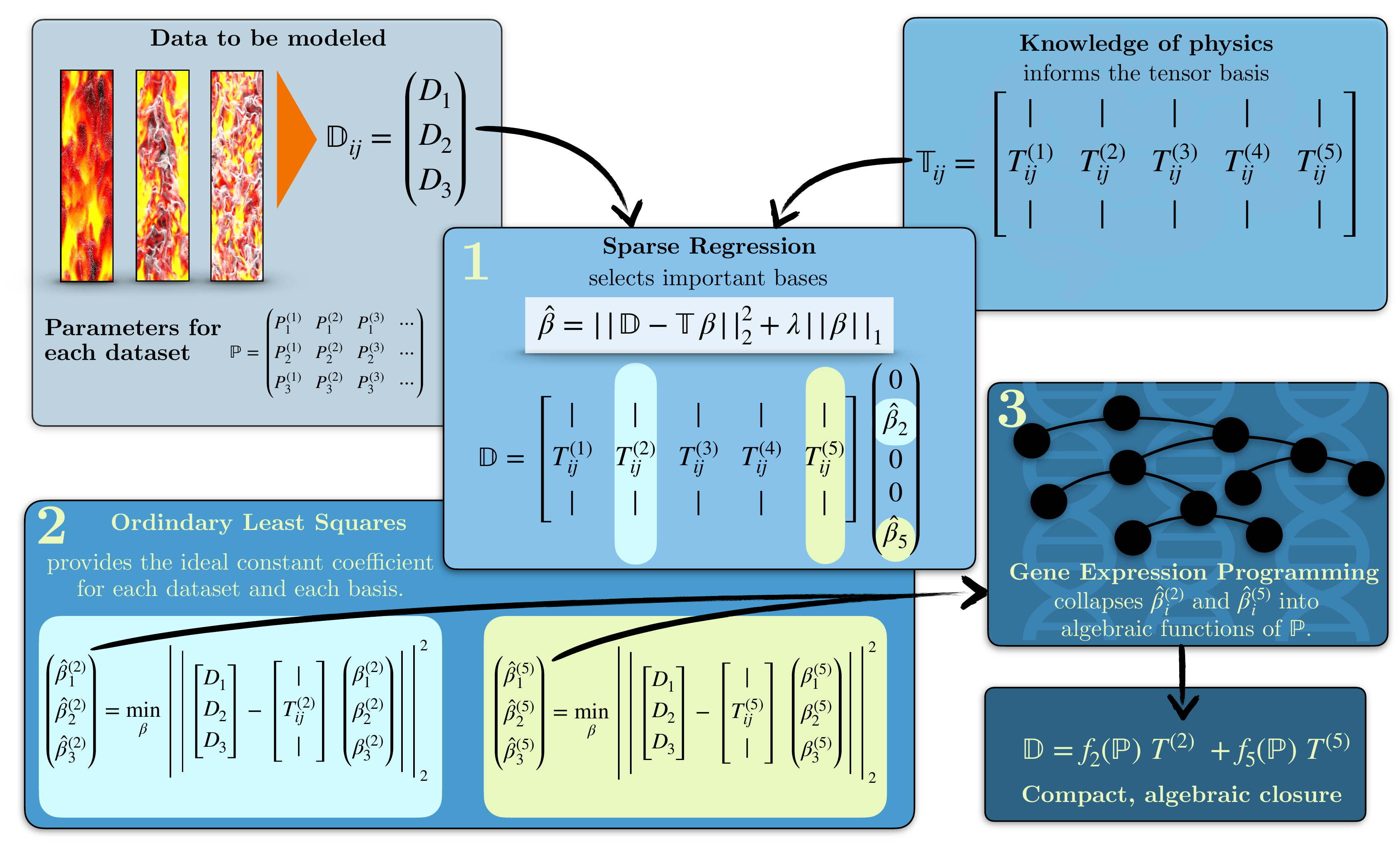}
    \caption{The modeling methodology has three steps: (1) Sparse regression identifies the important basis tensors, (2) OLS squares provides the ideal coefficients for each of the data sets for each of the identified bases and (3) GEP collapses the ideal coefficients for each case into a compact, algebraic closure.}
    \label{fig:modelflow}
\end{figure}


The method can be summarized by three distinct modeling steps, as shown in Fig.~\ref{fig:modelflow}, and outlined here for data spanning $s$ unique conditions in the parameter space (in the context of multiphase, turbulent flows, these parameters might include solids volume fraction, Reynolds number, etc.):
\begin{itemize}
    \item[1.] Use sparse regression to identify the basis tensors required to describe physics by optimizing    
    \begin{equation}
        \hat{\beta} = \min_{\beta} \vert \vert \mathbb{D} - \mathbb{T} \beta \vert \vert_2^2 + \lambda \vert \vert \beta \vert \vert_1  
    \end{equation}
    and assuming constant coefficients. Each base associated with a nonzero coefficient in $\hat{\beta}$ is deemed to be `essential' and is retained in the final model. The surviving bases are then condensed into a subset of $\mathbb{T}$, denoted $T^{\subset}$. 
    \item[2.] For each of the $s$ conditions, compute the ideal constant coefficients associated with the $p$ essential bases, using Ordinary Least Squares (OLS):  
    \begin{equation} \hat{\beta}^{\subset}_s = \min_{\hat{\beta}^{\subset}_s}\vert \vert {D}_s - {T}_s^{\subset} \beta^{\subset}_s\vert \vert^2_2, 
    \end{equation} 
    where $\hat{\beta}_s^{\subset}$ is of size $p\times 1$, $D_s$ is size $q\times 1$, where $q$ is the amount of data in the dataset (e.g. the number of time steps) and $T^{\subset}_s$ is size $q \times p$. Note that both $T^{\subset}$ and $D$ require tensorial data to be reorganized as vertical vectors \cite[see][for details]{Beetham2020}. After this process has been done for all $s$ conditions, concatenate each of the $\hat{\beta}^{\subset}_s$ vectors into a matrix of size $p\times s$. 
    \item[3.] Finally, provide each $p$-th row of $\hat{\beta}^{\subset}$ and matrix of parameters, $\mathbb{P}$, associated with the $s$ conditions as input to the GEP algorithm. The resulting functional model for $\hat{\beta}^{\subset}$ effectively collapses the vector of discrete values for $\hat{\beta}^{\subset}$ to a continuous, closed form with algebraic dependence on system parameters.
\end{itemize}

This modeling flow is illustrated in Fig.~\ref{fig:modelflow}, where $s=3$ and $p=2$ for demonstration purposes.

\section{Case study description}
\label{sec:config} 
Multiphase flows span large parameter spaces, making modeling challenging. Thus, we use a simple gas--solid flow in which two-way coupling between the phases drives the underlying turbulence as a case study to evaluate the effectiveness of the proposed modeling framework. 
In this configurations, rigid spherical particles are suspended in an unbounded (triply periodic) domain containing an initially quiescent gas. As particles settle under the influence of gravity, they spontaneously form clusters. Due to momentum exchange between phases, particles entrain the fluid, generating turbulence therein.  A frame of reference with the fluid phase is considered, such that the mean streamwise fluid velocity is null. Key non-dimensional numbers that characterize the system include the Reynolds number, the Archimedes number, defined as
\begin{equation}
{\rm Ar} =  (\rho_p/\rho_f -1)d_p^3 g/\nu_f^2.
\end{equation} 
Alternatively, a Froude number can be introduced to characterize the balance between gravitational and inertial forces, defined as $\Fr= \tau_p^2 g/ d_p$, where $\tau_p = {\rho_p d_p^2}/{(18 \rho_f \nu_f)}$ is the particle response time. 

The mean particle-phase volume fraction is varied from $0.001\le\langle \varepsilon_p \rangle \le0.05$ and gravity is varied from $0.8\le g\le8.0$. Here, angled brackets denote an average in all three spatial dimensions and time. Due to the large density ratios under consideration, the mean mass loading, $\varphi=\langle \varepsilon_p \rangle \rho_p/\left( \langle \varepsilon_f \rangle \rho_f \right)$, ranges from $\mathcal{O}(10)$--$\mathcal{O}(10^2)$, and consequently two-way coupling between the phases is important. A large enough domain with a sufficiently large number of particles is needed to observe clustering. To enable simulations on this scale, we use an Eulerian--Lagrangian approach \citep{Capecelatro2013}. Fluid equations are solved on a staggered grid with second-order spatial accuracy and advanced in time with second-order accuracy using the semi-implicit Crank--Nicolson scheme.  Particles are tracked individually in a Lagrangian frame of reference and integrated using a second-order Runge--Kutta method. Particle data is projected to the Eulerian mesh using a Gaussian filter described in \citet{Capecelatro2013}. 

Derivation of the \textit{single-phase} RANS equations is done by directly averaging the Navier--Stokes equations. Derivation of the \textit{multiphase} RANS equations, however, will retain additional physics if averaging is performed on the volume-filtered Navier-Stokes equations~\cite{fox2014}. Volume fraction weighted averages, or phase averaging (PA), analogous to Favre averaging of variable density flows, has previously been derived \citep{capecelatro2015}. For the relatively simple configuration used here, which is homogeneous in all spatial directions, statistically stationary in time and symmetric in the counter-gravity direction, the transport equations for the fluid-phase Reynolds stresses can be reduced to two unique, non-zero components. In the streamwise direction this equation is given as
\begin{align}
\label{eq:PA_ufuf}
\frac{1}{2} \frac{\partial \langle u_f^{\prime \prime \prime 2}\rangle_f}{\partial t } = &\underbrace{\frac{1}{\rho_f} \left \langle p_f \frac{\partial  u_f^{\prime \prime \prime} }{\partial x}\right \rangle }_{\text{\scriptsize{pressure strain (PS)}}} -  \underbrace{\frac{1}{\rho_f} \left \langle \sigma_{f,1i} \frac{\partial u_f^{\prime \prime \prime} }{\partial x_i}\right \rangle }_{\text{\scriptsize{viscous dissipation (VD)}}} + \underbrace{ \frac{\varphi}{\tau_p^{\star}}\left( \langle u_f^{\prime \prime \prime} u_p^{\prime\prime} \rangle_p - \langle u_f^{\prime \prime \prime 2}\rangle_p \right)}_{\text{\scriptsize{drag exchange (DE)}}} + \nonumber \\
&\underbrace{\frac{\varphi}{\tau_p^{\star}} \langle u_f^{\prime \prime \prime}\rangle_p \langle u_p\rangle_p }_{\text{\scriptsize{drag production (DP)}}} + \underbrace{\frac{\varphi}{\rho_p} \left \langle u_f^{\prime\prime\prime}\frac{\partial p^{\prime}_f}{\partial x} \right \rangle_p }_{\text{\scriptsize{pressure exchange (PE)}}} - \underbrace{\frac{\varphi}{\rho_p} \left \langle  u_f^{\prime \prime \prime} \frac{\partial \sigma^{\prime}_{f,1i}}{\partial x_i} \right \rangle_p }_{\text{\scriptsize{viscous exchange (VE)}}},
\end{align}
Where $u_p$ is the particle-phase velocity in an Eulerian frame of reference. Here, $\langle (\cdot) \rangle_p = \langle \varepsilon_p (\cdot) \rangle/\langle \varepsilon_p \rangle$. Fluctuations about PA terms are denoted with a double prime.  In a similar fashion, the PA operator in the fluid phase is defined as $\langle (\cdot) \rangle_f = \langle \varepsilon_f (\cdot) \rangle/\langle \varepsilon_f \rangle$. Fluctuations about the PA fluid velocity are given by $\bm{u}_f^{\prime \prime \prime} = \bm{u}_f(\bm{x},t) - \langle \bm{u}_f \rangle_f$. With this, the fluid-phase TKE is given by $k_f = \langle \bm{u}_f^{\prime \prime \prime} \cdot  \bm{u}_f^{\prime \prime \prime} \rangle_f/2$.

It is notable that all the terms appearing on the right hand side of \eqref{eq:PA_ufuf} are unclosed and require modeling. This work has already been carried out using sparse regression exclusively \citep{Beetham2021}. Here, we select the drag production term to demonstrate the present methodology. This term is chosen due to its importance in this class of flows. In the absence of mean shear, it is the only source of fluid-phase TKE. Additionally, all components of the drag production tensor are identically zero, except for the contribution in the gravity aligned direction. This condition often presents challenges for modeling. 


\section{Results and discussion} 
\label{sec:modeling}
We now demonstrate the modeling methodology presented in Sec.~\ref{sec:method} on the multiphase case study summarized in Sec.~\ref{sec:config}, focusing on the drag production term, $\mathcal{R}^{\rm{DP}}$, in particular. Here, we follow the three modeling steps as outlined previously. 

In the first step, we  conduct modeling of drag production using sparse regression with embedded invariance and the assumption of constant coefficients to inform the bases that comprise the reduced set, ${T}^{\subset}$. The model consisting of the reduced basis tensors is given as
\begin{equation} 
\mathcal{R}^{\rm{DP}} = \beta_1 \mathbb{I} + \beta_2 \hat{\mathbb{U}}_r,
\end{equation} 
where $\hat{\mathbb{U}}_r$ is a tensor formulated using the mean slip velocity between the phase, $\mathbb{I}$ is the identity tensor and the coefficients, $\beta_1$ and $\beta_2$, have functional dependency upon configuration parameters that are unknown and cannot be informed by physics-based reasoning.

 Next, we evaluate the ideal \emph{constant} coefficients for each unique configuration studied, by conducting OLS and allowing the coefficients, $\beta_1$ and $\beta_2$ to take on unique values for each configuration. In other words, the values of $\beta_1$ and $\beta_2$ associated with the case for $\langle \varepsilon_p \rangle = 0.001$ and $g=0.8$ need not be the same as the values for $\langle \varepsilon_p \rangle = 0.05$ and $g=2.4$. 
 
 As described in Sec.~\ref{sec:method}, the ideal coefficients are arranged into two vectors: one for each of the basis tensors, $\mathbb{I}$ and $\hat{\mathbb{U}}_r$. Each vector of ideal coefficients is used as input, along with the associated parameters and invariants, to the GEP algorithm \citep{Searson2009}. Here, the GEP algorithm selects models that reduce the $R^2$ between the ideal coefficient values and the candidate models, which are all functions of the parameters and invariants. This effectively collapses the vector of ideal coefficients to a single, compact algebraic expression. 

The resultant model learned from this methodology is given as 
\begin{align}
\mathcal{R}^{\rm{DP}} = &\left( 0.258 \varphi + (0.03 \varphi)^3 + 1.9 \frac{\langle \varepsilon_p \rangle}{\mathcal{S}^{(2)}} \right)\mathbb{I} + \\ \nonumber
   &\left(1.9 \varphi - 5.8 \varphi^{1/2}\right)\hat{\mathbb{U}}_r 
\end{align}
where $\mathcal{S}^{(2)}$ is a principal invariant, defined as ${\rm tr}\left(\hat{\mathbb{U}}_r \hat{\mathbb{R}}_f \hat{\mathbb{R}}_p^2\right)$, and the basis tensor, $\hat{\mathbb{U}}_r$ is defined by the normalized slip tensor. This slip tensor is given as the outer product of the mean slip velocity, $\left(\langle \bm{u}_p \rangle_p - \langle \bm{u}_f \rangle_f \right) \otimes \left(\langle \bm{u}_p \rangle_p - \langle \bm{u}_f \rangle_f \right)$. The two other basis tensors, $\hat{\mathbb{R}}_f$ and $\hat{\mathbb{R}}_p$, are the anisotropic stress tensors associated with the fluid and particle phase, respectively. In terms of solution variables, the mean phase velocities are solved by associated momentum equations and the Reynolds stresses are informed by transport equations in the multiphase RANS equations \cite[see][]{Beetham2021}.
This model has an error of 0.012, where the error is defined as
\begin{equation}
\epsilon = \frac{\vert \vert \mathbb{D} - \mathbb{T}\hat{\beta} \vert \vert^2_2}{\vert \vert \mathbb{D} \vert \vert^2_2}.
\end{equation}
This is comparable performance to the model learned using sparse regression exclusively ($\epsilon = 0.013$), however, the proposed method does not require a manual selection of trial functions for the coefficients, thus making it far more efficient from a modeling perspective. 

As a counter argument to the blended modeling approach, we also allowed the GEP algorithm to learn the full model (i.e., the mean values of drag production, all 24 basis tensors, the principal invariants and configuration parameters from the Euler--Lagrange simulations were provided to the GEP algorithm as input). The learned model is given as 
\begin{align}
    \mathcal{R}^{\rm{DP}} = &24.4 \hat{\mathbb{U}}_r + 30.4 e^{-\hat{\mathbb{R}}_p} - 1.42 \left(\hat{\mathbb{U}}_r\hat{\mathbb{R}}_p + (\hat{\mathbb{U}}_r\hat{\mathbb{R}}_p)^{\rm T}\right)^{1/2} + \\ \nonumber 
    &1.41\times 10^5 \left( \hat{\mathbb{U}}_r^2 \hat{\mathbb{R}}_f + (\hat{\mathbb{U}}_r^2 \hat{\mathbb{R}}_f )^{\rm T}\right)^2 \langle \varepsilon_p \rangle^2 - 30.4,
\end{align}
with associated error 0.13 (an order of magnitude higher than the blended or sparse regression only approach).  This degradation in performance can be attributed to the fact that the model now depends upon $\hat{\mathbb{R}}_p$ and $\hat{\mathbb{R}}_f$, the particle and fluid-phase Reynolds stress tensors, in addition to $\hat{\mathbb{U}}_r$. On a fundamental level, since the drag production is a gravity-based phenomenon (i.e., TKE is generated \emph{solely} due to the presence of gravity in this configuration), we can anticipate that $\hat{\mathbb{U}}_r$ would be the predominant tensor from the basis for describing the physics. Additionally, since the Reynolds stresses contain nonzero diagonal entries, including these terms makes it difficult to drive the cross stream directions of the drag production model to zero. Finally, and perhaps most importantly, GEP does not enforce the relation that resultant model be linear with respect to the basis tensors. The stipulation of linearity in the basis tensors is critical for ensuring form invariance in the resultant model and for ensuring a physics-based description of the data, as described by \eqref{eq:linear}. These results suggest that sparse regression and GEP are both needed in order to select a minimal set of tensors to describe physics and automate the complex dependencies of the coefficients when physical intuition cannot guide this process. 

\section{Acknowledgements} 
This material is based upon work supported by the National Science Foundation (NSF CAREER, CBET-1846054). The computing resources and assistance provided by the staff of Advanced Research Computing at the University of Michigan, Ann Arbor is greatly appreciated. Additionally, this work used the Extreme Science and Engineering Discovery Environment (XSEDE), which is supported by National Science Foundation grant number ACI-1548562 \citep{XSEDE}. 

\bibliographystyle{plainnat}
\bibliography{IUTAM}

\begin{thebibliography}{28}
\providecommand{\natexlab}[1]{#1}
\providecommand{\url}[1]{\texttt{#1}}
\expandafter\ifx\csname urlstyle\endcsname\relax
  \providecommand{\doi}[1]{doi: #1}\else
  \providecommand{\doi}{doi: \begingroup \urlstyle{rm}\Url}\fi

\bibitem[Beetham and Capecelatro(2020)]{Beetham2020}
S.~Beetham and J.~Capecelatro.
\newblock Formulating turbulence closures using sparse regression with embedded
  form invariance.
\newblock \emph{Physical Review Fluids}, 5:\penalty0 084611, 2020.

\bibitem[Beetham et~al.(2021)Beetham, Fox, and Capecelatro]{Beetham2021}
S.~Beetham, R.~O. Fox, and J.~Capecelatro.
\newblock Sparse identification of multiphase turbulence closures for coupled
  fluid-particle flows.
\newblock \emph{Journal of Fluid Mechanics}, 914, A11, 2021.

\bibitem[Bode et~al.(2019)Bode, Gauding, Kleinheinz, and Pitsch]{Bode2019}
M.~Bode, M.~Gauding, K.~Kleinheinz, and H.~Pitsch.
\newblock Deep learning at scale for subgrid modeling in turbulent flows:
  regression and reconstruction.
\newblock \emph{arXiv:1910.00928v1}, 2019.

\bibitem[Brunton et~al.(2016)Brunton, Proctor, and Kutz]{ML_2016Brunton}
S.~L. Brunton, J.~L. Proctor, and J.~N. Kutz.
\newblock Discovering governing equations from data by sparse identification of
  nonlinear dynamical systems.
\newblock \emph{Proceedings of the National Academy of Sciences},
  113(15):\penalty0 3932--3937, 2016.

\bibitem[Capecelatro and Desjardins(2013)]{Capecelatro2013}
J.~Capecelatro and O.~Desjardins.
\newblock An {E}uler--{L}agrange strategy for simulating particle-laden flows.
\newblock \emph{Journal of Computational Physics}, 238:\penalty0 1--31, 2013.

\bibitem[Capecelatro et~al.(2014)Capecelatro, Desjardins, and
  Fox]{capecelatro2014cit}
J.~Capecelatro, O.~Desjardins, and R.~O. Fox.
\newblock Numerical study of collisional particle dynamics in cluster-induced
  turbulence.
\newblock \emph{Journal of Fluid Mechanics}, 747:\penalty0 R2 1--13, 2014.

\bibitem[Capecelatro et~al.(2015)Capecelatro, Desjardins, and
  Fox]{capecelatro2015}
J.~Capecelatro, O.~Desjardins, and R.~O. Fox.
\newblock On fluid-particle dynamics in fully-developed cluster-induced
  turbulence.
\newblock \emph{Journal of Fluid Mechanics}, 780:\penalty0 578--635, 2015.

\bibitem[Dominique et~al.(2021)Dominique, Christophe, Schram, and
  Sandberg]{dominique2021inferring}
Joachim Dominique, Julien Christophe, Christophe Schram, and Richard~D
  Sandberg.
\newblock Inferring empirical wall pressure spectral models with {G}ene
  {E}xpression {P}rogramming.
\newblock \emph{Journal of Sound and Vibration}, 506:\penalty0 116162, 2021.

\bibitem[Duraisamy and Durbin(2014)]{ML_2014Duraisamy_transition}
K.~Duraisamy and P.~A. Durbin.
\newblock Transition modeling using data driven approaches.
\newblock \emph{Proceedings of the Summer Program}, page 427, 2014.

\bibitem[Duraisamy et~al.(2015)Duraisamy, Zhang, and
  Singh]{ML_2014Duraisamy_new}
K.~Duraisamy, Z.~J. Zhang, and A.~P. Singh.
\newblock New approaches in turbulence and transition modeling using
  data-driven techniques.
\newblock \emph{53rd AIAA Aerospace Sciences Meeting}, page 1284, 2015.

\bibitem[Duraisamy et~al.(2019)Duraisamy, Iaccarino, and
  Xiao]{duraisamy2019turbulence}
Karthik Duraisamy, Gianluca Iaccarino, and Heng Xiao.
\newblock Turbulence modeling in the age of data.
\newblock \emph{Annual Review of Fluid Mechanics}, 51:\penalty0 357--377, 2019.

\bibitem[Fox(2014)]{fox2014}
R.~O. Fox.
\newblock On multiphase turbulence models for collisional fluid--particle
  flows.
\newblock \emph{Journal of Fluid Mechanics}, 742:\penalty0 368--424, 2014.

\bibitem[Ling et~al.(2016)Ling, Kurzawski, and Templeton]{Ling2016}
J.~Ling, A.~Kurzawski, and J.~Templeton.
\newblock Reynolds averaged turbulence modeling using deep neural networks with
  embedded invariance.
\newblock \emph{Journal of Fluid Mechanics}, 807:\penalty0 155--166, 2016.

\bibitem[Liu and Fang(2019)]{Liu2019}
W.~Liu and J.~Fang.
\newblock Iterative framework of machine-learning based turbulence modeling for
  {R}eynolds-averaged {N}avier-{S}tokes simulations.
\newblock \emph{arXiv:1910.01232v1}, 2019.

\bibitem[Lu(2010)]{ML_2010Lu}
C.~Lu.
\newblock Artificial neural network for behavior learning from meso-scale
  simulations, application to multi-scale multimaterial flows.
\newblock \emph{PhD thesis}, 2010.

\bibitem[Ma et~al.(2016)Ma, Lu, and Tryggvason]{ML_2016Ma}
M.~Ma, J.~Lu, and G.~Tryggvason.
\newblock Using statistical learning to close two-fluid multiphase flow
  equations for bubbly flows in vertical channels.
\newblock \emph{International Journal of Multiphase Flow}, 85:\penalty0
  336--347, 2016.

\bibitem[Milano and Koumoutsakos(2002)]{ML_2002Milano}
M.~Milano and P.~Koumoutsakos.
\newblock Neural network modeling for near wall turbulent flow.
\newblock \emph{Journal of Computational Physics}, 182:\penalty0 1--26, 2002.

\bibitem[Pope(2000)]{PopeText}
S.~B. Pope.
\newblock Turbulent flows.
\newblock \emph{Cambridge University Press}, 2000.

\bibitem[Rajabi and Kavianpour(2012)]{ML_2012Rajabi}
E.~Rajabi and M.~R. Kavianpour.
\newblock Intelligent prediction of turbulent flow over backward-facing step
  using direct numerical simulation data.
\newblock \emph{Engineering Applications of Computational Fluid Mechanics},
  6(4):\penalty0 490--503, 2012.

\bibitem[Reissmann et~al.(2021)Reissmann, Hasslberger, Sandberg, and
  Klein]{reissmann2021application}
Maximilian Reissmann, Josef Hasslberger, Richard~D Sandberg, and Markus Klein.
\newblock Application of {G}ene {E}xpression {P}rogramming to a-posteriori
  {LES} modeling of a {T}aylor {G}reen vortex.
\newblock \emph{Journal of Computational Physics}, 424:\penalty0 109859, 2021.

\bibitem[Samadianfard(2012)]{samadianfard2012gene}
Saeed Samadianfard.
\newblock Gene expression programming analysis of implicit colebrook--white
  equation in turbulent flow friction factor calculation.
\newblock \emph{Journal of Petroleum Science and Engineering}, 92:\penalty0
  48--55, 2012.

\bibitem[Schmelzer et~al.(2020)Schmelzer, Dwight, and
  Cinnella]{schmelzer2020discovery}
M.~Schmelzer, R.~P. Dwight, and P.~Cinnella.
\newblock Discovery of algebraic {R}eynolds-stress models using sparse symbolic
  regression.
\newblock \emph{Flow, Turbulence and Combustion}, 104\penalty0 (2):\penalty0
  579--603, 2020.

\bibitem[Searson(2009)]{Searson2009}
D.~Searson.
\newblock Gptips: Genetic programming \& symbolic regression for {MATLAB}.
\newblock \emph{http://gptips.sourceforge.net}, 2009.

\bibitem[Spencer and Rivlin(1958)]{Spencer1958}
A.~J.~M. Spencer and R.~S. Rivlin.
\newblock The theory of matrix polynomials and its application to the mechanics
  of isotropic continua.
\newblock \emph{Archive for Rational Mechanics and Analysis}, 2:\penalty0
  309--336, 1958.

\bibitem[Towns et~al.(2014)Towns, Cockerill, Dahan, Foster, Gaither, Grimshaw,
  Hazlewood, Lathrop, Lifka, Peterson, Roskies, Scott, and
  Wilkins-Diehr]{XSEDE}
J.~Towns, T.~Cockerill, M.~Dahan, I.~Foster, K.~Gaither, A.~Grimshaw,
  V.~Hazlewood, S.~Lathrop, D.~Lifka, G.~D. Peterson, R.~Roskies, J.~R. Scott,
  and N.~Wilkins-Diehr.
\newblock Xsede: Accelerating scientific discovery.
\newblock \emph{Computing in Science \& Engineering}, 16:\penalty0 62--74,
  2014.

\bibitem[Tracey et~al.(2015)Tracey, Duraisamy, and Alonso]{ML_2015Tracey}
B.~Tracey, K.~Duraisamy, and J.~J. Alonso.
\newblock A machine learning strategy to assist turbulence model development.
\newblock \emph{AIAA Paper}, 1287, 2015.

\bibitem[Weatheritt and Sandberg(2016)]{weatheritt2016novel}
Jack Weatheritt and Richard Sandberg.
\newblock A novel evolutionary algorithm applied to algebraic modifications of
  the {RANS} stress--strain relationship.
\newblock \emph{Journal of Computational Physics}, 325:\penalty0 22--37, 2016.

\bibitem[Zhao et~al.(2020)Zhao, Akolekar, Weatheritt, Michelassi, and
  Sandberg]{zhao2020rans}
Yaomin Zhao, Harshal~D Akolekar, Jack Weatheritt, Vittorio Michelassi, and
  Richard~D Sandberg.
\newblock {RANS} turbulence model development using cfd-driven machine
  learning.
\newblock \emph{Journal of Computational Physics}, 411:\penalty0 109413, 2020.

\end{thebibliography}

\end{document}